\documentclass[12pt,preprint]{aastex}

\begin{document}

\title{The Radio-Loud Fraction of Quasars is a Strong Function of Redshift and 
  Optical Luminosity}
\author{Linhua Jiang\altaffilmark{1}, Xiaohui Fan\altaffilmark{1},
  \v{Z}eljko Ivezi\'{c}\altaffilmark{2}, Gordon T. Richards\altaffilmark{3,4},
  Donald P. Schneider\altaffilmark{5}, Michael A. Strauss\altaffilmark{3},
  and Brandon C. Kelly\altaffilmark{1}}
\altaffiltext{1}{Steward Observatory, University of Arizona, 933 North Cherry 
  Avenue, Tucson, AZ 85721}
\altaffiltext{2}{Department of Astronomy, University of Washington, Box
  351580, Seattle, WA 98195}
\altaffiltext{3}{Department of Astrophysical Sciences, Peyton Hall, Princeton,
  NJ 08544}
\altaffiltext{4}{Department of Physics and Astronomy, The Johns Hopkins
  University, 3400 North Charles Street, Baltimore, MD 21218}
\altaffiltext{5}{Department of Astronomy and Astrophysics, Pennsylvania
  State University, 525 Davey Laboratory, University Park, PA 16802}

\begin{abstract}

Using a sample of optically-selected quasars from the Sloan Digital Sky 
Survey, we have determined the radio-loud fraction (RLF) of quasars as a 
function of redshift and optical luminosity. The sample contains more than 
30,000 objects and spans a redshift range of $0<z\leqslant5$ and a luminosity 
range of $-30\leqslant M_{i}<-22$. We use both the radio-to-optical flux ratio 
($R$ parameter) and radio luminosity to define radio-loud quasars. After 
breaking the correlation between redshift and luminosity due to the 
flux-limited nature of the sample, we find that the RLF of quasars decreases 
with increasing redshift and decreasing luminosity. The relation can be 
described in the form of log(RLF/(1--RLF)) 
$=b_0+b_z \log(1+z)+b_M (M_{2500}+26)$, where $M_{2500}$ is the absolute 
magnitude 
at rest-frame 2500 {\AA}, and $b_z$, $b_M<0$. When using $R>10$ to define 
radio-loud quasars, we find that $b_0=-0.132\pm0.116$, $b_z=-2.052\pm0.261$,
and $b_M=-0.183\pm0.025$. The RLF at $z=0.5$ declines from 24.3\% to 5.6\% as
luminosity decreases from $M_{2500}=-26$ to $M_{2500}=-22$, and the RLF at 
$M_{2500}=-26$ declines from 24.3\% to 4.1\% as redshift increases 
from 0.5 to 3, suggesting that the RLF is a strong function of both redshift 
and luminosity. We also examine the impact of flux-related selection effects 
on the RLF determination using a series of tests, and find that the dependence
of the RLF on redshift and luminosity is highly likely to be physical, and the 
selection effects we considered are not responsible for the dependence.

\end{abstract}


\section{Introduction}

Although quasars were first discovered by their radio emission
\citep[e.g.][]{mat63,sch63}, it was soon found that the majority of quasars 
were radio-quiet \citep[e.g.][]{san65}. Quasars are often classified into two 
broad categories, radio-loud and radio-quiet, based on their radio properties.
There is mounting evidence that the distribution of radio-to-optical flux 
ratio for optically-selected quasars is bimodal 
\citep[e.g.][]{kel89,mil90,vis92,ive02}, although the existence of the 
bimodality has been questioned \citep[e.g.][but see also \citet{ive04a} for a 
response]{cir03}. Radio-loud and radio-quiet quasars are probably 
powered by similar physical mechanisms \citep[e.g.][]{bar89,urr95}, and their 
radio properties are also correlated with host galaxy properties, central 
black hole masses, black hole spins, and accretion rates 
\citep[e.g.][]{bau95,urr95,bes05}. Radio-loud quasars are likely to reside in 
more massive galaxies \citep[e.g.][]{pea86,bes05}, and harbor more massive 
central black holes \citep[e.g.][]{lao00,lac01,mcl04}, than do radio-quiet 
quasars.

Roughly 10\%$-$20\% of all quasars are radio-loud 
\citep[e.g.][]{kel89,urr95,ive02}. However, the radio-loud fraction (RLF) of 
quasars may depend on redshift and optical luminosity. Some studies have 
found that the RLF tends to drop with increasing redshift 
\citep[e.g.][]{pea86,mil90,vis92,sch92} and decreasing luminosity 
\citep[e.g.][]{pad93,gol99,cir03}, or evolves non-monotonically with redshift 
and luminosity \citep[e.g.][]{hoo95,bis97}, while others showed that the RLF 
does not differ significantly with redshift \citep[e.g.][]{gol99,ste00,cir03} 
or luminosity \citep[e.g.][]{bis97,ste00}.

From a sample of 4472 quasars from the Sloan Digital Sky Survey 
\citep[SDSS;][]{yor00}, \citet{ive02} found that the RLF is independent of 
both redshift and optical luminosity when using marginal distributions of the 
whole sample; however, they noted that the approximate degeneracy between 
redshift and luminosity in the SDSS flux-limited sample may cause individual 
trends in redshift and luminosity to appear to cancel. 
By stacking the images of the Faint Images of the Radio Sky at Twenty-cm
survey \citep[FIRST;][]{bec95}, \citet{whi06} were able to probe the radio
sky into nanoJansky regime. They found that the median radio loudness of
SDSS-selected quasars is a declining function with optical luminosity.
After correcting for this effect, they claimed that the median radio loudness
is independent of redshift. In this paper we use a 
sample of more than 30,000 optically-selected quasars from the SDSS, and break 
the redshift-luminosity dependence to study the evolution of the RLF. We will 
find that there are indeed strong trends in redshift and luminosity, and that 
they do in fact roughly cancel in the marginal distributions.

In $\S$2 of this paper, we present our quasar sample from the SDSS. In $\S$3 
we derive the RLF of quasars as a function of redshift and optical 
luminosity. We examine the effects of K corrections and sample incompleteness
in $\S$4, and we give the discussion and summary in $\S$5 and $\S$6,
respectively. Throughout the paper we use a $\Lambda$-dominated flat cosmology 
with H$_0$ = 70 km s$^{-1}$ Mpc$^{-1}$, $\Omega_{m}$ = 0.3, and 
$\Omega_{\Lambda}$ = 0.7 \citep[e.g.][]{spe06}.

\section{The SDSS quasar sample}

The SDSS \citep{yor00} is an imaging and spectroscopic survey of the sky 
using a dedicated wide-field 2.5m telescope \citep{gun06}. The imaging is 
carried out in five broad bands, $ugriz$, spanning the range from 3000 to 
10,000 {\AA} \citep{fuk96,gun98}. From the resulting catalogs of objects,
quasar candidates \citep{ric02} are selected for spectroscopic follow-up.
Spectroscopy is performed using a pair of double spectrographs with coverage
from 3800 to 9200 {\AA}, and a resolution $\lambda/\Delta \lambda$ of
roughly 2000. The SDSS quasar survey spectroscopically targets quasars 
with $i<19.1$ at low redshift ($z\leqslant3$) and $i<20.2$ at high redshift 
($z\geqslant 3$). The low-redshift selection is performed in $ugri$ color 
space, and the high-redshift selection is performed in $griz$ color space. 
In addition to the optical selection, a SDSS object is also considered to be 
a primary quasar candidate if it is an optical point source located within 
$2\farcs0$ of a FIRST radio source. All SDSS magnitudes mentioned in this 
paper have been corrected for Galactic extinction using the maps of 
\citet{sch98}.

The sample we used is from the SDSS Data Release Three 
\citep[DR3;][]{aba05}. The quasar catalog of the DR3 consists of 46,420 
objects with luminosities larger than $M_{i}=-22$ \citep{sch05}. The area 
covered by the catalog is about 3732 deg$^2$. We reject 4683 objects that are 
not covered by the FIRST survey, 
and we only use the quasars which were selected on their 
optical colors (i.e., the quasars with one or more of the following target 
selection flags: QSO\_HIZ, QSO\_CAP and QSO\_SKIRT; see Richards et al. 2002)
to avoid the bias introduced by the FIRST 
radio selection. The final sample consists of 31,835 optically-selected 
quasars from the SDSS DR3 catalog, and covers a redshift range of 
$0<z\leqslant5$ and a luminosity range of $-30\leqslant M_{i}<-22$.

To include both core-dominated (hereafter FR1) and lobe-dominated (hereafter
FR2) quasars, we match our sample to the FIRST catalog \citep{whi97}
with a matching radius 30$\arcsec$.
For the quasars that have only one radio source within 30$\arcsec$, we match
them again to the FIRST catalog within 5$\arcsec$ and classify the matched 
ones as FR1 quasars. The quasars that have multiple entries within
30$\arcsec$ are classified as FR2 quasars. The sample contains 2566 
FIRST-detected quasars, including 1944 FR1 quasars and 622 FR2 quasars.

We use the integrated flux density ($f_{int}$) in the FIRST catalog to 
describe the 20 cm radio emission. The total radio flux density of each FR2 
quasar is determined using all of the radio components 
within 30$\arcsec$. We note that we have excluded those FR2 quasars whose
separations between lobes are greater than 1$\arcmin$. In fact, FR2 quasars
represent a small fraction of the SDSS DR3 catalog, and FR2 quasars
with diameters greater than 1$\arcmin$ are even rarer \citep{dev06}. 
These numbers are too small to affect the statistics below. Therefore
we do not use more sophisticated procedures \citep[e.g.][]{ive02,dev06} to
select FR2 quasars. 

\section{RLF of quasars as a function of redshift and optical luminosity}

We define a radio-loud quasar based on its $R$ parameter, the rest-frame 
ratio of the flux density at 6 cm (5 GHz) to the flux density at 2500 {\AA} 
\citep[e.g.][]{sto92}. For a given quasar, we calculate its observed flux 
density $f_{6cm}$ at rest-frame 6 cm from $f_{int}$ (if detected) assuming 
a power-law slope of $-0.5$ \citep[e.g.][]{ive04b}; and we determine its 
observed flux density $f_{2500}$ at rest-frame 2500 {\AA} by fitting a model 
spectrum to the SDSS broadband photometry \citep{fan01,jia06,ric06}.
The model spectrum is a power-law continuum ($f_{\nu}=A{\nu}^{\alpha}$) plus 
a series of emission lines extracted from the quasar composite spectrum 
\citep{van01}. We integrate the model spectrum over the redshifted SDSS
bandpasses to compare with the observed magnitudes.
The parameters $\alpha$ and $A$ are 
determined by minimizing the differences between the model spectrum 
magnitudes $m^{model}$ and the SDSS photometry $m^{obs}$:
\begin{equation}
  {\chi}^2=\sum\left(\frac{m^{model}_i-m^{obs}_i}{\sigma^{obs}_i}\right)^2,
\end{equation}
where $\sigma^{obs}_{i}$ is the estimated SDSS photometry error in the 
$i^{\rm th}$ SDSS filter. We constrain $\alpha$ to be in the range 
$-1.1<\alpha<0.1$, and only use the bands that are not dominated by Lyman 
forest absorption systems. Finally $f_{2500}$ is computed from the power-law 
continuum using the best-fit values of $\alpha$ and $A$, and the radio 
loudness $R$ is obtained by
\begin{equation}
  R=f_{6cm}/f_{2500}.
  \label{eq:loudness}
\end{equation}
The absolute magnitude $M_{2500}$ at rest-frame 2500 {\AA} is calculated 
from $f_{2500}$.

The FIRST survey has a 5$\sigma$ peak flux density limit of about 1.0 mJy 
\citep{bec95}, although this limit is not perfectly uniform across the sky. 
For a quasar detected by FIRST, we 
determine the relevant limit directly from the FIRST catalog, while for 
a quasar undetected by FIRST, we measure the limit at the position of the 
nearest radio source (usually within $10\arcmin$). We find that the median 
value of the limits is 0.98 mJy, which has already included the effect of
``CLEAN bias'' \citep{bec95,whi97}. Only $\sim4$\% of the quasars have limits 
above 1.1 mJy, so we use 1.1 mJy as the FIRST detection limit for our sample. 

Many sources in the FIRST images are resolved. The resolution effect causes
FIRST to become more incomplete for extended objects near the detection limit
\citep{bec95,whi97}. Furthermore, FR2 quasars are more incomplete than FR1 
quasars for integrated flux densities. For example, a double-lobe radio 
source with two identical components suffers from incompleteness twice as 
high as a single-component source of the same total flux density. Figure 1
(provided by R. L. White, private communication) shows the FIRST completeness
as a function of integrated flux density. The completeness is computed using
the observed size distribution and rms values of integrated flux densities
from the FIRST survey for SDSS quasars, and has included all effects 
mentioned above. Quasars with $f_{int}>5$ mJy have a completeness fraction 
$g_{comp}\approx 1$ (100\%); while for a quasar with 1.1 mJy $<f_{int}<$ 5 
mJy, its $g_{comp}$ is measured from the curve. To correct for sample 
incompleteness, we use the weight of $1/g_{comp}$ when we calculate the 
numbers of radio-loud quasars.

When quasars with $R>10$ are defined as radio-loud 
\citep[e.g.][]{kel89}, FIRST is able to detect radio-loud quasars down to 
$i\approx 18.9$ based on Equation~\ref{eq:loudness}, the K corrections we 
applied and the FIRST detection limit of 1.1 mJy. The left panel of Figure 2 
shows the redshift and absolute magnitude distribution of our sample. 

In flux-limited surveys, redshift and luminosity are artificially correlated, 
making it difficult to separate the dependence of the RLF on redshift or 
luminosity. To break this degeneracy, we divide the $M_{2500}$--$z$ plane 
into small grids. RLFs in individual $M_{2500}$--$z$ grids are calculated and 
presented as squares in the right panel of Figure 2, where the square 
for each subsample is located at the median values of $M_{2500}$ and $z$ in 
that subsample. The RLF declines with 
increasing redshift and decreasing luminosity. One can see the trend more
clearly in Figure 3, in which we plot the RLF in three small redshift ranges
and three small magnitude ranges.

We assume a simple relation to model the RLF as a function of redshift and 
absolute magnitude,
\begin{equation}
  \log \left(\frac{RLF}{1-RLF} \right)=b_0+b_z \log(1+z)+b_M (M_{2500}+26),
  \label{eq:fracrl}
\end{equation}
where $b_0$, $b_z$ and $b_M$ are constants. We use the RLFs calculated from 
the grids that include more than one radio-loud quasar, and use median values
of $z$ and $M_{2500}$ in each grid. Statistical uncertainties are estimated
from Poisson statistics. The best fitting results found by regression fit are, 
$b_0=-0.132\pm0.116$, $b_z=-2.052\pm0.261$, and $b_M=-0.183\pm0.025$. This 
implies that when RLF $\ll1$, RLF $\propto (1+z)^{-2.052} L_{opt}^{0.458}$, 
where $L_{opt}$ is the optical luminosity. The $\chi^{2}$ of this fit is 52.8 
for 47 degrees of freedom (DoF), and the confidence levels of $b_z$ and $b_M$ 
are shown in Figure 4. The null hypothesis that $b_z=0$ and $b_M=0$ is 
rejected at $>5\sigma$ significance. The results 
are projected onto two-dimensional plots in the left panels of Figure 5, where 
filled circles are the RLFs calculated from individual grids in Figure 2 and 
dashed lines are the best fits. The upper panel shows the RLF as a function 
of redshift after correcting for the luminosity dependence. At $M_{2500}=-26$,
the RLF drops from $\sim$24.3\% to $\sim$4.1\% as the redshift increases from 
0.5 to 3. The lower panel shows the RLF as a function of luminosity after 
correcting for the redshift dependence. At $z=0.5$, the RLF decreases rapidly
from $\sim$24.3\% to $\sim$5.6\% as the luminosity decreases from 
$M_{2500}=-26$ to $M_{2500}=-22$. Therefore the RLF of quasars is a strong 
function of both redshift and optical luminosity.

To probe whether the trend seen in Equation~\ref{eq:fracrl} is related to the 
radio-loud criterion adopted, we define a radio-loud quasar if $R$ is greater 
than 30 instead of 10. In this case FIRST is able to detect radio-loud 
quasars down to $i\approx 20.0$. We calculate RLFs for the quasars with 
$i<20.0$ using the same method illustrated in Figure 2, and model the RLF 
with Equation~\ref{eq:fracrl}. The best fitting parameters are given in 
Table 1 and the confidence levels of $b_z$ and $b_M$ are shown in Figure 4.
Figure 5 shows the RLF as a function of redshift and luminosity. The relation
gives similar results for both radio-loud criteria. We note that the sample 
of $i<20.0$ at $z<3$ is not complete since some bins in $M_{2500}$--$z$ space
are not sampled by SDSS. However, this incompleteness does 
not bias our results because we are considering the RLF for optically-selected
quasars. Another definition of a radio-loud quasar is based on the radio 
luminosity of an object \citep[e.g.][]{pea86,mil90,hoo95,gol99}. As we do for 
the $R$-based RLF, we use two criteria to define radio-loud quasars: $L_r$ 
(luminosity density at rest-frame 6 cm) $>10^{32}$ ergs s$^{-1}$ Hz$^{-1}$ and 
$L_r>10^{32.5}$ ergs s$^{-1}$ Hz$^{-1}$. In the two cases, FIRST is able 
to detect radio-loud quasars up to $z\sim2.1$ and 3.5, respectively. We model 
the RLF using Equation~\ref{eq:fracrl} and repeat the analysis. The best
fitting results are shown in Table 1, and Figures 4 and 6. The RLF based on 
$L_r$ is correlated with $z$ and $M_{2500}$ in the same manner as the 
$R$-based criteria.

\section{Effects of the K corrections and sample incompleteness}

When applying the K corrections, we assumed that the slope of the radio 
continuum is $-0.5$ and used a model spectrum to determine the optical 
continuum slope. To investigate the effect of the K corrections, we performed 
several experiments. First, for a given quasar at $z$, we calculated its 
$f_{2500}$ and $M_{2500}$ from the magnitude in the SDSS band whose effective 
wavelength is closest to $2500(1+z)$ {\AA} assuming a slope of $-0.5$ 
(Test 1). As before, we corrected for the contribution from emission lines. 
Because the SDSS $ugriz$ photometry covers a wavelength range of 3000 to 
10,000 \AA, the K corrections for $z<3$ require no extrapolation. 
Second, we assumed two extreme cases for the radio and optical slopes: in 
Test 2, we took the optical slope to be 0.0 and the radio slope as $-1.0$, 
while in Test 3, the optical slope was $-1.0$ and the radio slope was 0.0.
The results of the fit to Equation~\ref{eq:fracrl} are listed in Table 1. The 
values of $b_z$ and $b_M$ recalculated under these tests differ by less
than 2$\sigma$ from the original values, and the null hypothesis that 
$b_z=0$ and $b_M=0$ is rejected at $>5\sigma$ significance in all these
tests. Therefore the effect of the K corrections on our conclusions is small.

We investigated the reliability of the relation described by
Equation~\ref{eq:fracrl} for different definitions of radio loudness and for
luminosities in different optical bands. For example, we defined the $R$
parameter as the rest-frame ratio of the flux density at 6 cm to the flux 
density at 4400 {\AA} \citep[e.g.][]{kel89} instead of 2500 {\AA}, and we 
repeated the analysis in $\S$3 (Test 4). In Test 5, we determined the RLF as 
a function of $z$ and $M_i$ (instead of $M_{2500}$). $M_i$ is the absolute 
magnitude in the
rest-frame $i$ band, and was calculated using the method described in $\S$3.
The results are listed in Table 1. In these cases 
the RLF is still strongly dependent on redshift and luminosity, and the 
relation described by Equation~\ref{eq:fracrl} is not sensitive to the 
details of how radio loudness is defined.

When examining the dependence of the RLF on $M_{2500}$ and $z$, we note that
contours of constant RLF in Figure 2 roughly coincide with contours of  
constant apparent magnitude. This is illustrated in Figure 7(a), which shows 
that for different redshift bins, the relation between the RLF and 
{\em apparent i magnitude} is independent of redshift at $i>17.5$. 
The RLF does decrease with redshift at $i<17.5$, although with large error 
bars. This ``conspiracy'' of strong dependence of the RLF on apparent 
magnitude raises the concern that our results have been affected by 
flux-dependent selection effects. In this paper we are considering the RLF
for optically-selected quasars, and the SDSS color selection is highly
complete at $z<2.2$ \citep{ric06}, so optical selection effects are not 
likely to seriously affect 
the RLF determination. However, the SDSS quasar selection becomes 
increasingly incomplete for objects with very red intrinsic colors 
\citep[$\alpha<-1.5$,][]{fan01,ric02}, especially at high redshift.
\citet{whi06} found a strong correlation between radio loudness and optical
color using the SDSS sample. We reproduce this dependence in the upper panel 
of Figure 8. The RLF rises with increasing $\Delta(g-i)$, which is the 
difference between the observed $g-i$ and the median $g-i$ color of quasars
at that redshift, following \citet{hop04}. To examine 
whether this RLF-color dependence affects the relation in 
Equation~\ref{eq:fracrl}, we divide the quasar sample into several 
$\Delta(g-i)$ bins, and calculate the RLF as a function of $M_{2500}$ and
$z$ for each bin of {\em intrinsic} quasar colors.
We find that although the average RLF increases toward redder continuum, 
the RLF is still a strong function of $M_{2500}$ and $z$ within each bin, 
similar to the relation shown in Equation~\ref{eq:fracrl}. The lower panel of 
Figure 8 gives an example for the bin of $-0.1<\Delta(g-i)<0.1$. Note that 
more than 80\% of the quasars in the sample lie in the range of 
$-0.3<\Delta(g-i)<0.2$, within which the RLF-color
relation is relatively flat. Therefore, although we can not determine 
accurately the RLF evolution for the reddest few percent of quasars where
SDSS is incomplete, the strong correlation between the RLF and both 
redshift and luminosity is not strongly affected by the RLF-color relation 
over the color range in which the SDSS selection is essentially complete.

In order to examine radio selection effects, we performed the following tests.
\begin{enumerate}
\item Did we miss FR1 quasars due to the 5$\arcsec$ radio catalog matching? 
We used a 10$\arcsec$ matching instead, and found that the number of FR1 
quasars increases by only 3.7\%. We also found that these additional 
sources increase the RLF by similar factors at both high and low redshift 
and both high and low luminosity, and thus have little effect on $b_z$ or 
$b_M$.
\item Did we measure the radio fluxes of FR2 quasars correctly? We compared
FIRST with the NRAO VLA Sky Survey \citep[NVSS;][]{con98}, which has a 
resolution of 45$\arcsec$. The FIRST and NVSS fluxes of most FR2 quasars are 
in good agreement. FIRST has a resolution of 5$\arcsec$ and may overresolve 
radio sources larger than about 10$\arcsec$. These sources are rare and very 
bright in radio (usually $>50$ mJy), well above the radio-loud division in 
our analysis.
\item Were there quasars detected by NVSS but not by FIRST? We matched our 
sample to the NVSS catalog within 15$\arcsec$. We found that about 6\% of the 
matched quasars were not detected by FIRST, and 80\% of these additional 
sources are FR1 sources with offsets more than 5$\arcsec$ from the SDSS 
positions. They increase the RLF by similar factors at both high and 
low redshift and both high and low luminosity, and thus do not significantly
change the trend in the RLF.
\item How did the incompleteness of FIRST at the detection limit affect our 
results? In Section 3 we use 1.1 mJy as the FIRST detection limit, and we
correct for the incompleteness caused by the resolution effect. Here
we set two tests to examine the incompleteness near the FIRST limit. In Test 6
we use a limit of 1.5 mJy (the corresponding $i\sim18.6$) to determine the 
RLF for $R>10$; while in Test 7 we use a limit of 3 mJy (the corresponding 
$i\sim18.9$) to calculate the RLF for $R>30$. Note that the FIRST completeness 
measured from the completeness curve in Figure 1 is 75\%
at 1.5 mJy and 95\% at 3 mJy. The results of the tests are given 
in Table 1, and they show the similar trend of the RLF on redshift and 
optical luminosity as we obtained above.

\end{enumerate}
Therefore, based on these tests we conclude that the strong dependence of the 
RLF on $M_{2500}$ and $z$ is highly likely to be physical, and the radio 
incompleteness and selection effects we have considered are not responsible 
for the dependence, though we cannot rule out other unexplored selection 
effects.

\section{Discussion}

Most of the previous studies of the RLF are based on samples of tens to 
hundreds of quasars. Considering that the RLF is only $\sim10$\% on average, 
these samples are not large enough to study the two-dimensional distribution 
of the RLF on $M_{2500}$ and $z$. This makes it difficult to uncover the 
relation of Equation~\ref{eq:fracrl} using only the marginal distribution of 
the RLF. We calculate the marginal RLF as a function of $M_{2500}$ and $z$
for our sample, 
which is shown in Figures 7(b) and 7(c). One can see the marginal RLF is 
roughly independent of both $M_{2500}$ and $z$, because the dependence on 
$M_{2500}$ and $z$ roughly cancel out due to the $M_{2500}$--$z$ degeneracy. 
This result is in quantitative agreement with the marginal RLF derived by 
\citet{ive02}.

It has been suggested that high-redshift quasars show little difference
in their rest-frame UV/optical and X-ray properties from those of low-redshift
quasars. Their emission-line 
strengths and UV continuum shapes are very similar to those of low-redshift 
quasars \citep[e.g.][]{bar03,pen03,fan04}, the emission line ratios indicate 
solar or supersolar metallicity in emission-line regions as found in 
low-redshift quasars \citep[e.g.][]{ham99,die03,mai03}, and the 
optical-to-X-ray flux ratios and X-ray continuum shapes show little evolution 
with redshift \citep[e.g.][]{str05,ste06,she06}. These measurements suggest 
that most quasar properties are not sensitive to the cosmic age. However,
Figure 2 shows that the RLF evolves strongly with redshift, thus the evolution 
of the RLF places important constraints on models of quasar 
evolution and the radio emission mechanism. 

Equation~\ref{eq:fracrl} implies a strong correlation between the RLF and 
optical luminosity. Using a large sample of low-redshift ($0.03<z<0.3$) 
AGNs, \citet{bes05} find that radio-loud AGNs tend to reside in old, massive 
galaxies, and that the fraction of radio-loud
AGNs is a strong function of stellar mass or central black hole mass (e.g., 
the fraction increases from zero at a stellar mass of $10^{10}$M$_{\sun}$ to 
30\% at a stellar mass of $5\times10^{11}$M$_{\sun}$). Assuming that optical
luminosity is roughly proportional to black hole mass \citep{pet04}, their
radio-loud fraction of AGNs is also a strong function of optical luminosity. 
This is in qualitative agreement with Equation~\ref{eq:fracrl}, although our
quasars are more luminous than their low-redshift AGNs. 

Equation~\ref{eq:fracrl} also shows that the RLF is a strong function of 
redshift. By stacking FIRST images of SDSS-selected quasars, \citet{whi06} 
recently found that the median $R$ is a declining function
with optical luminosity. After correcting for this effect, they claimed that
the median $R$ is independent on redshift, which seems inconsistent with
our result that the RLF is a strong negative function of redshift. However,
the median $R$ and the RLF are not identical. The median $R$ is determined by 
the majority of quasars with low $R$ values (i.e. radio-quiet quasars); while 
the RLF is the fraction of quasars exceeding a threshold in $R$, and therefore 
corresponds to the behavior of the small fraction of quasars with high $R$ 
values (i.e. radio-loud quasars).
There are two natural ways to interpret the evolution of the RLF
with redshift \citep[e.g.][]{pea86}: (1) This may be due to the cosmological
evolution of quasar radio properties, such as $R$ and $L_r$. For instance,
a decreasing $R$ results in a decreasing RLF for increasing redshift. (2) This
could be simply caused by the density evolution of different populations of
quasars (e.g. radio-loud and radio-quiet quasars). The results of 
\citet{whi06} may have ruled out the first explanation and leave the second 
one: the different density evolution behaviors for the two classes of 
quasars. For instance, there are more radio-loud quasars at low redshift, but 
the fraction of radio-loud quasars is small, so they do not change the 
median $R$, which is still dominated by radio-quiet quasars.
This claim is based on stacked FIRST images. To distinguish between 
the two explanations, one needs to determine the radio luminosity function
of quasars in different redshift ranges, including the radio-quiet population, 
going to radio fluxes much fainter than those probed by FIRST.
Deep surveys such as the Cosmic 
Evolution Survey \citep{sch04} that cover a wide redshift range and reach
low luminosity in both optical and radio wavelengths are needed to 
interpret the evolution of the RLF.

\section{Summary}

In this paper we use a sample of more than 30,000 optically-selected quasars 
from SDSS to determine the RLF of quasars as a function of redshift and 
optical luminosity. The sample covers a large range of redshift and 
luminosity. We study the RLF using different criteria to define radio-loud 
quasars. After breaking the degeneracy between redshift and luminosity, we 
find that the RLF is a strong function of both redshift and optical 
luminosity: the RLF decreases rapidly with increasing redshift and decreasing 
luminosity. The relation can be described by a simple model, given by 
Equation~\ref{eq:fracrl}. We have done a series of tests to examine the 
impact of flux-related selection effects, and find that the dependence of the 
RLF on redshift and luminosity is highly likely to be physical.

The RLF is one of a few quasar properties that
strongly evolve with redshift, so the evolution of the RLF places important 
constraints on models of quasar evolution and the radio emission mechanism.
By comparing our results with the behavior of the median $R$ derived from
stacked FIRST images, we find that the evolution of the RLF with redshift
could be explained by the different density evolution for radio-loud 
and radio-quiet quasars. 
Substantially deeper wide-angle radio surveys which obtain fluxes 
for the radio-quiet population are needed to fully understand the physical 
nature of this evolution.

\acknowledgments

We acknowledge support from NSF grant AST-0307384, a Sloan Research Fellowship
and a Packard Fellowship for Science and Engineering (L.J., X.F.), and 
NSF grant AST-0307409 (M.A.S). We thank R. L. White and R. H. Becker for
helpful discussions.

Funding for the SDSS and SDSS-II has been provided by the Alfred P. Sloan
Foundation, the Participating Institutions, the National Science Foundation,
the U.S. Department of Energy, the National Aeronautics and Space
Administration, the Japanese Monbukagakusho, the Max Planck Society, and the
Higher Education Funding Council for England.
The SDSS Web Site is http://www.sdss.org/.
The SDSS is managed by the Astrophysical Research Consortium for the
Participating Institutions. The Participating Institutions are the American
Museum of Natural History, Astrophysical Institute Potsdam, University of
Basel, Cambridge University, Case Western Reserve University, University of
Chicago, Drexel University, Fermilab, the Institute for Advanced Study, the
Japan Participation Group, Johns Hopkins University, the Joint Institute for
Nuclear Astrophysics, the Kavli Institute for Particle Astrophysics and
Cosmology, the Korean Scientist Group, the Chinese Academy of Sciences
(LAMOST), Los Alamos National Laboratory, the Max-Planck-Institute for
Astronomy (MPIA), the Max-Planck-Institute for Astrophysics (MPA), New Mexico
State University, Ohio State University, University of Pittsburgh, University
of Portsmouth, Princeton University, the United States Naval Observatory, and
the University of Washington.

\clearpage
\begin{deluxetable}{lcccccc}
\rotate
\tablecaption{Results of fits to Equation~\ref{eq:fracrl}}
\tablewidth{0pt}
\tablehead{
\colhead{Sample} & \colhead{$\chi^{2}$} & \colhead{DoF} & \colhead{$b_0$} & 
\colhead{$b_z$}  & \colhead{$b_M$} & \colhead{Cov($b_z$,$b_M$)\tablenotemark{a}}}
\startdata
$R>10$          & 52.8 & 47 & $-0.132\pm0.116$ & $-2.052\pm0.261$ & $-0.183\pm0.025$ & 0.0059 \\
$R>30$          & 50.7 & 45 & $-0.218\pm0.110$ & $-2.096\pm0.240$ & $-0.203\pm0.023$ & 0.0050 \\
$\log L_r>32$   & 30.3 & 35 & $-0.053\pm0.122$ & $-2.214\pm0.277$ & $-0.307\pm0.024$ & 0.0055 \\
$\log L_r>32.5$ & 45.2 & 41 & $-0.216\pm0.118$ & $-2.088\pm0.260$ & $-0.333\pm0.025$ & 0.0055 \\
Test 1 ($R>10$) & 62.1 & 48 & $-0.328\pm0.115$ & $-1.639\pm0.259$ & $-0.141\pm0.024$ & 0.0058 \\
Test 2 ($R>10$) & 64.4 & 46 & $-0.292\pm0.116$ & $-1.693\pm0.259$ & $-0.165\pm0.024$ & 0.0058 \\
Test 3 ($R>10$) & 57.0 & 46 & $-0.308\pm0.114$ & $-1.722\pm0.261$ & $-0.123\pm0.024$ & 0.0058 \\
Test 4 ($R>10$) & 53.6 & 46 & $-0.142\pm0.120$ & $-2.115\pm0.268$ & $-0.194\pm0.025$ & 0.0063 \\
Test 5 ($R>10$) & 78.9 & 46 & $0.120\pm0.102$  & $-2.924\pm0.258$ & $-0.254\pm0.022$ & 0.0054 \\
Test 6 ($R>10$) & 51.9 & 41 & $-0.104\pm0.148$ & $-2.137\pm0.347$ & $-0.185\pm0.032$ & 0.0106 \\
Test 7 ($R>30$) & 49.2 & 45 & $-0.213\pm0.128$ & $-2.115\pm0.286$ & $-0.202\pm0.027$ & 0.0071 \\
\enddata
\tablenotetext{a}{Covariance between $b_z$ and $b_M$.}
\end{deluxetable}

\clearpage
\begin{figure}
\plotone{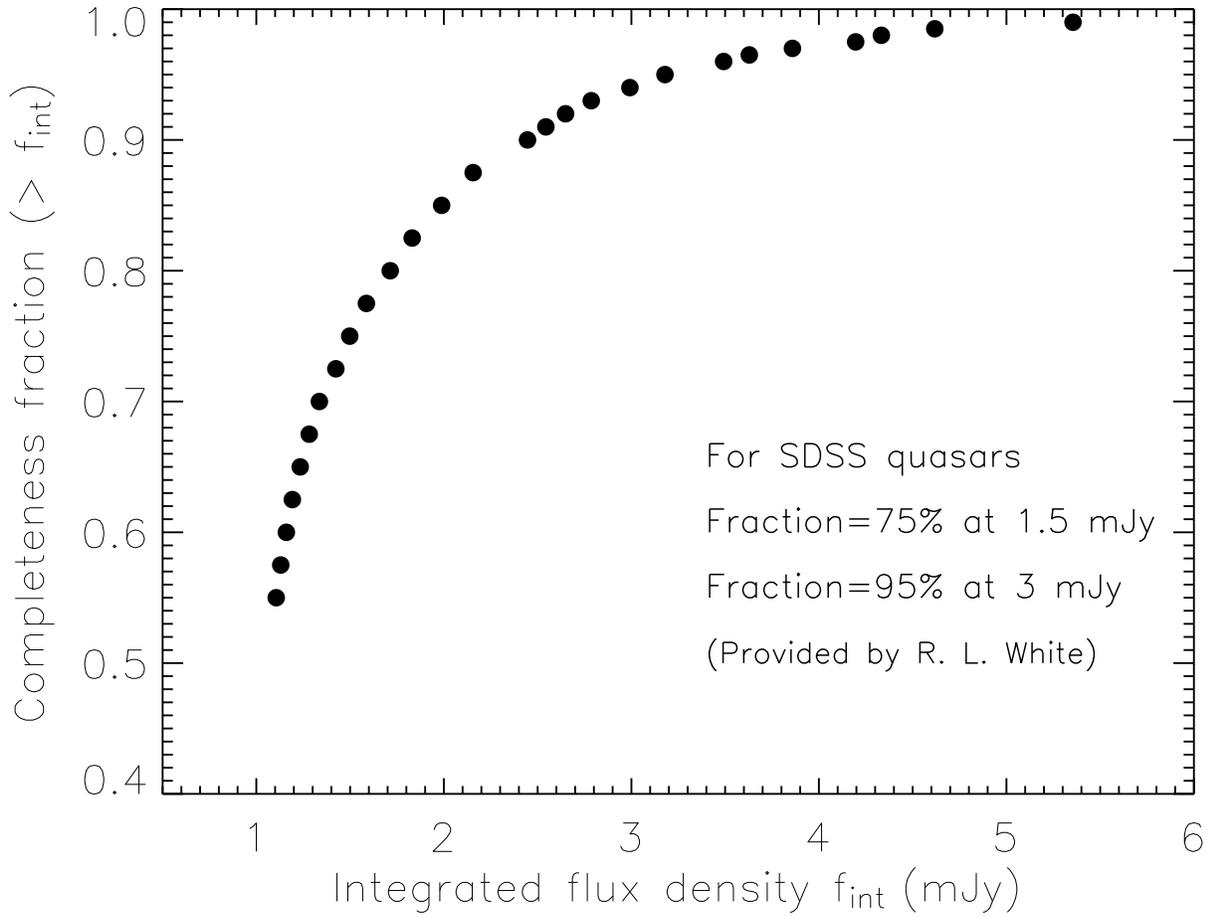}
\caption{The FIRST completeness as a function of integrated flux density
(provided by R. L. White, private communication). The completeness is 
computed using the observed size distribution of SDSS quasars and rms values 
of integrated flux densities from the FIRST survey.}
\end{figure}

\clearpage
\begin{figure}
\plotone{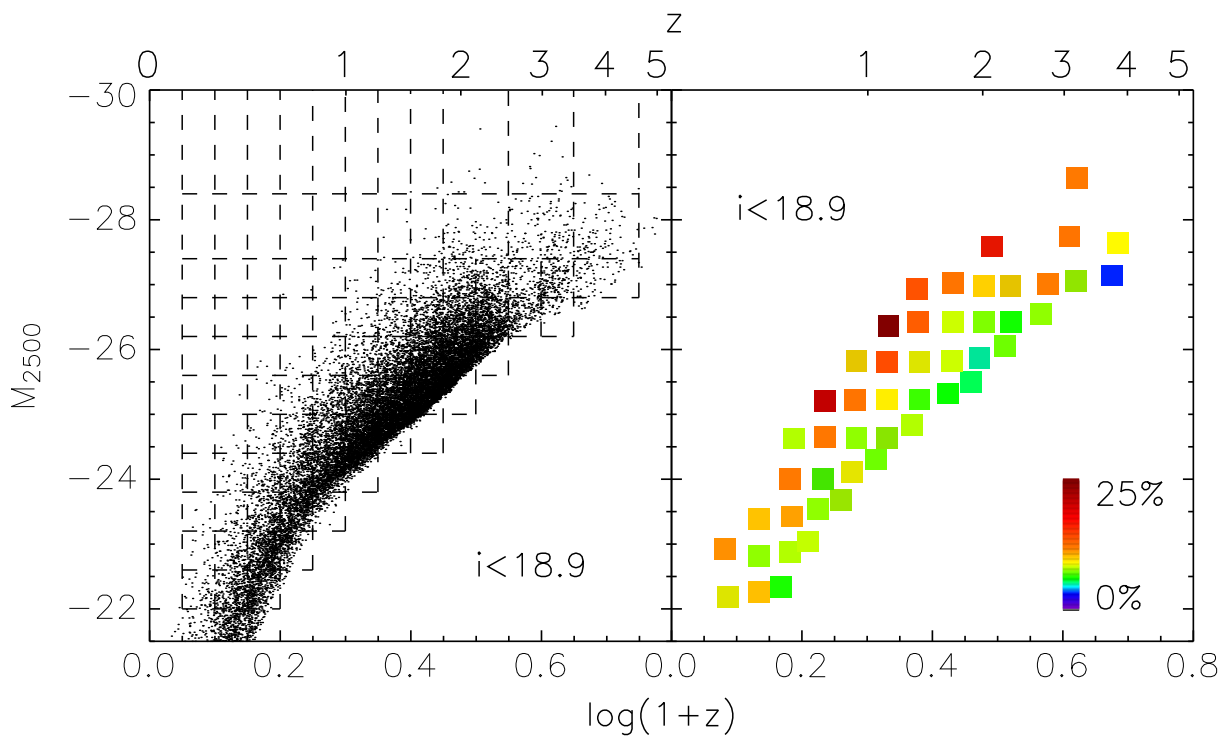}
\caption{Left panel: Redshift and absolute magnitude distribution for 20,473 
quasars with $i<18.9$ in our sample. The $M_{2500}$--$z$ plane is divided into 
small grids to break the redshift-luminosity dependence. RLFs of quasars are 
calculated in individual grids. Right panel: The $R$-based RLFs in individual
$M_{2500}$--$z$ bins. The square for each subsample is positioned at the 
median values of $M_{2500}$ and $z$ in that subsample. The RLF of quasars 
declines with increasing redshift and decreasing luminosity.}
\end{figure}

\clearpage
\begin{figure}
\plotone{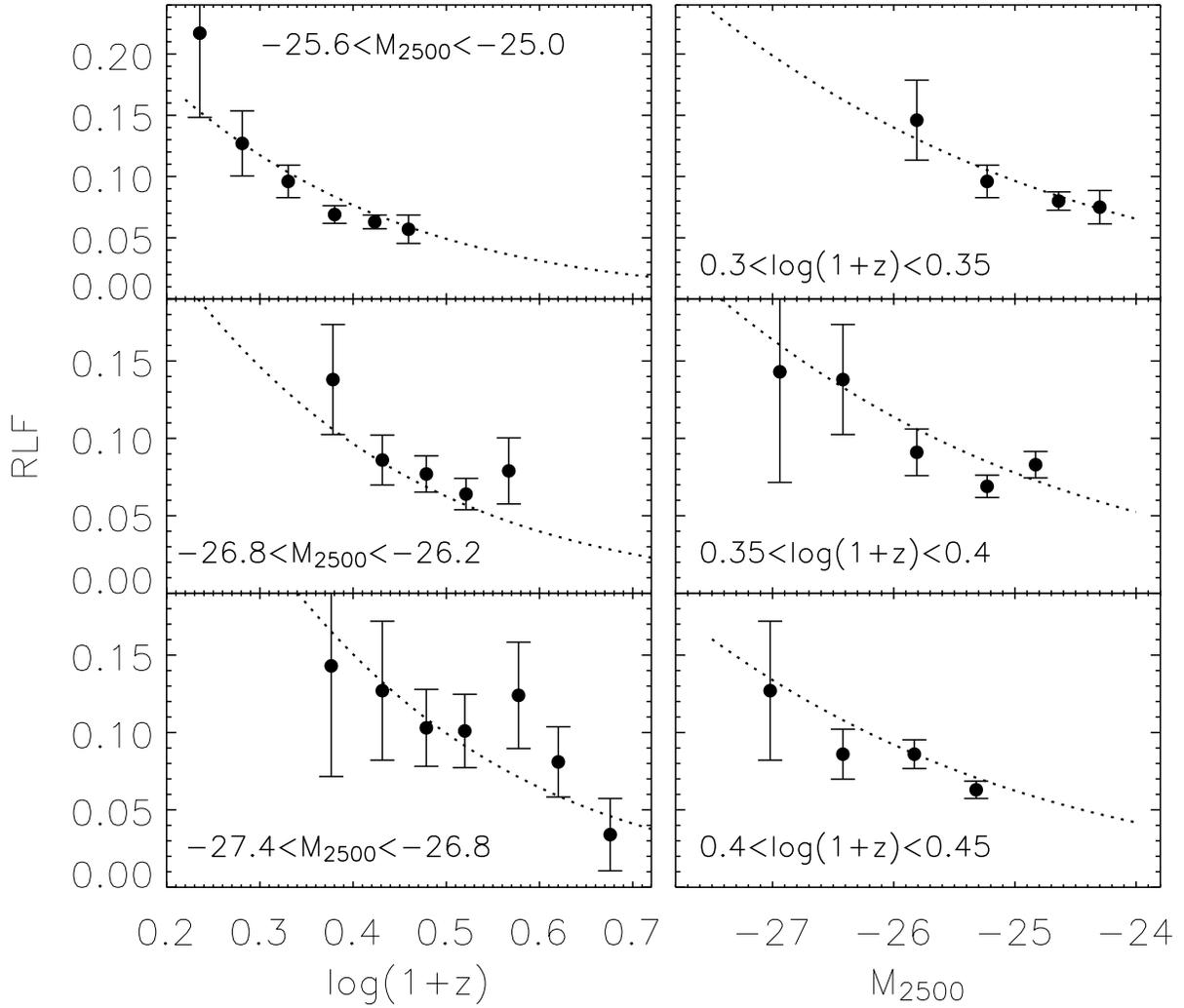}
\caption{
RLF in three small redshift ranges and three small magnitude ranges.
Dotted lines are the best model fits.}
\end{figure}                                                                            
\clearpage
\begin{figure}
\plotone{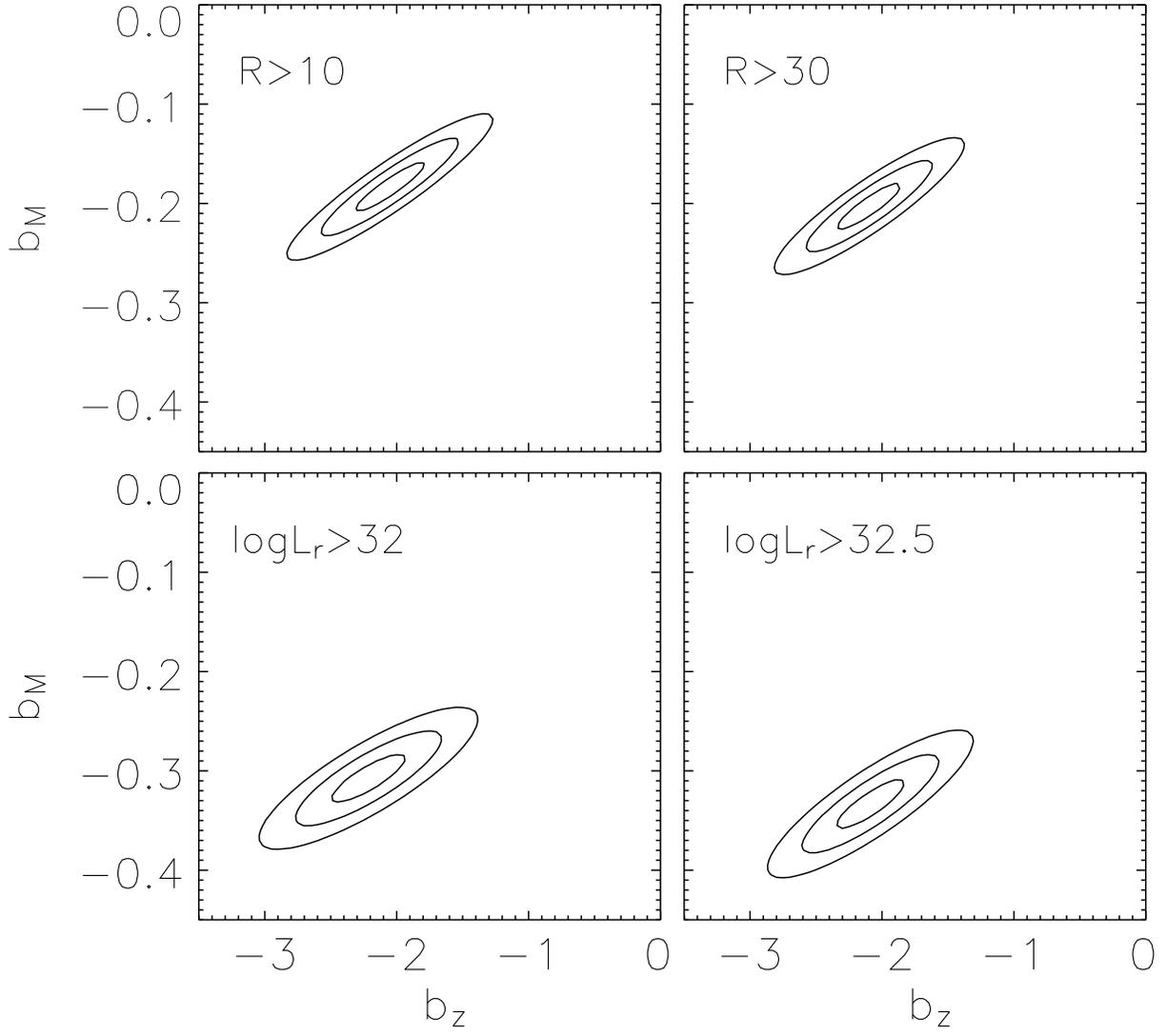}
\caption{
$1\sigma$, $2\sigma$ and $3\sigma$ confidence regions for $b_z$ vs. $b_M$.}
\end{figure}

\clearpage
\begin{figure}
\epsscale{0.8}
\plotone{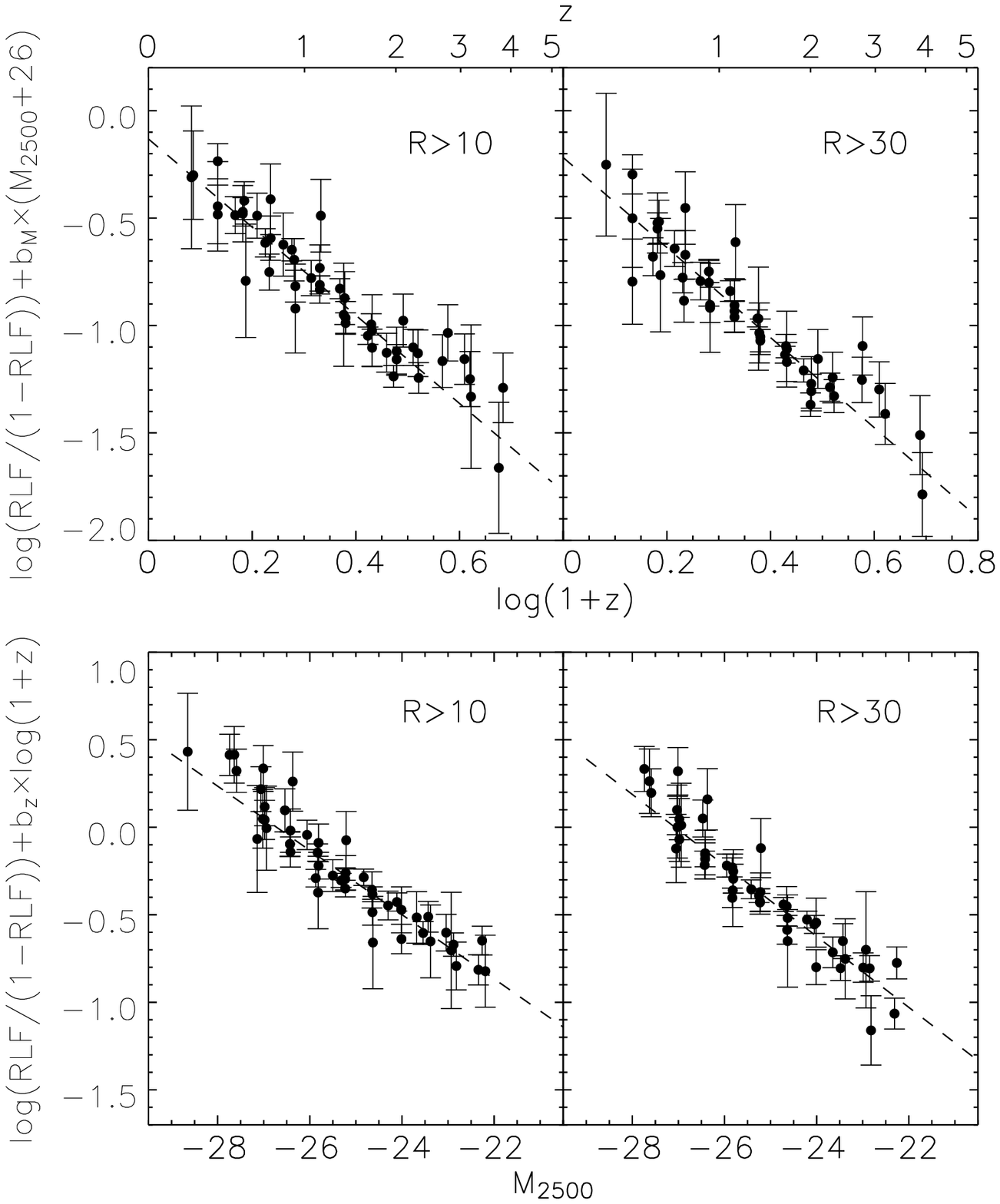}
\caption{The $R$-based RLF as a function of $z$ and $M_{2500}$. The upper 
panels show the RLF as a function of redshift after correcting for the 
luminosity dependence and the lower panels show the RLF as a function of 
luminosity after correcting for the redshift dependence. Filled circles are 
RLFs calculated from individual $M_{2500}$--$z$ grids and dashed lines are 
the best model fits. Poisson errors are also given in the figure.}
\end{figure}

\clearpage
\begin{figure}
\plotone{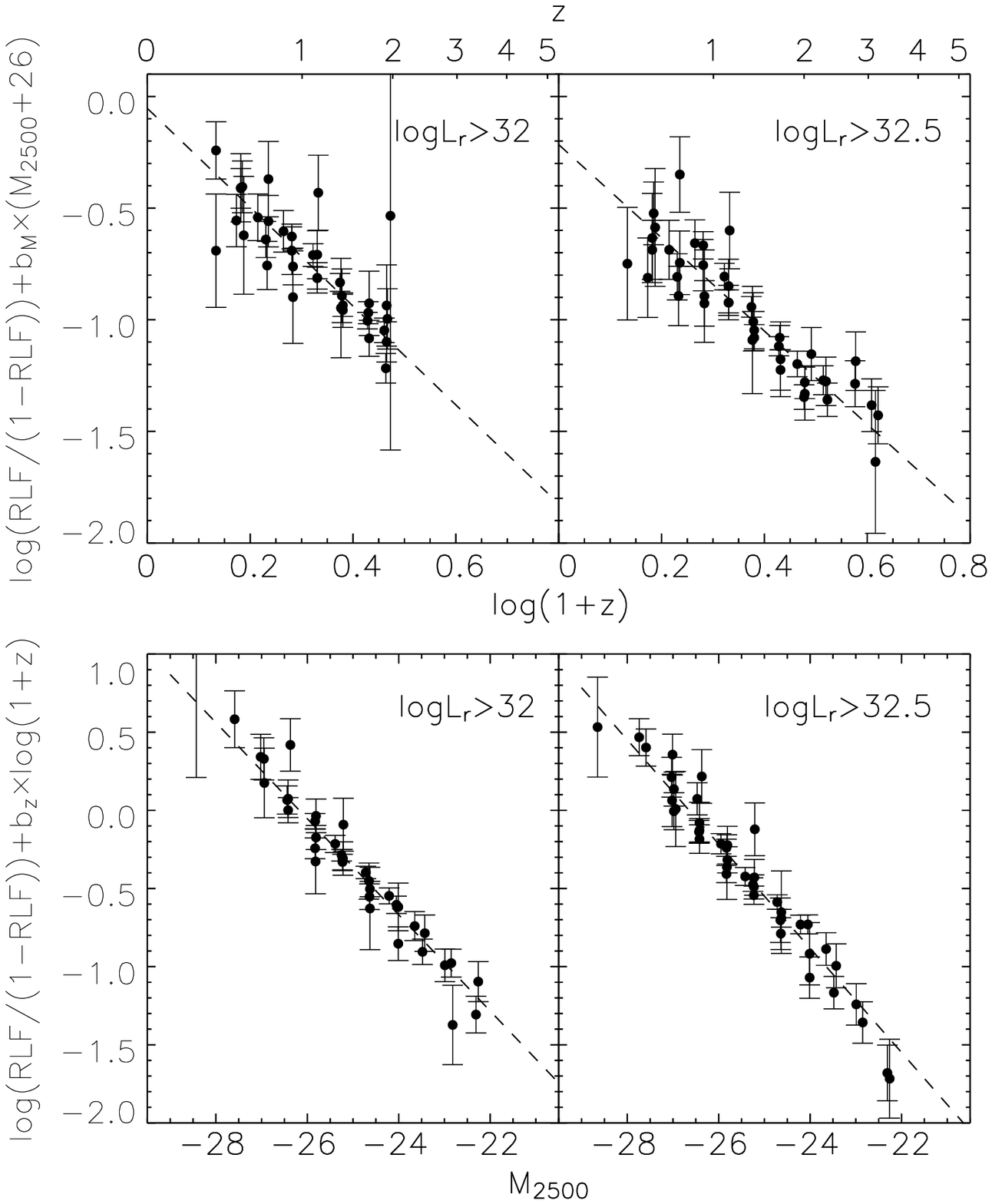}
\caption{The $L_r$-based RLF as a function of $z$ and $M_{2500}$. The upper 
panels show the RLF as a function of redshift after correcting for the 
luminosity dependence and the lower panels show the RLF as a function of 
luminosity after correcting for the redshift dependence. Filled circles are 
RLFs calculated from individual $M_{2500}$--$z$ grids and dashed lines are 
the best model fits. Poisson errors are also given in the figure.}
\end{figure}

\clearpage
\begin{figure}
\plotone{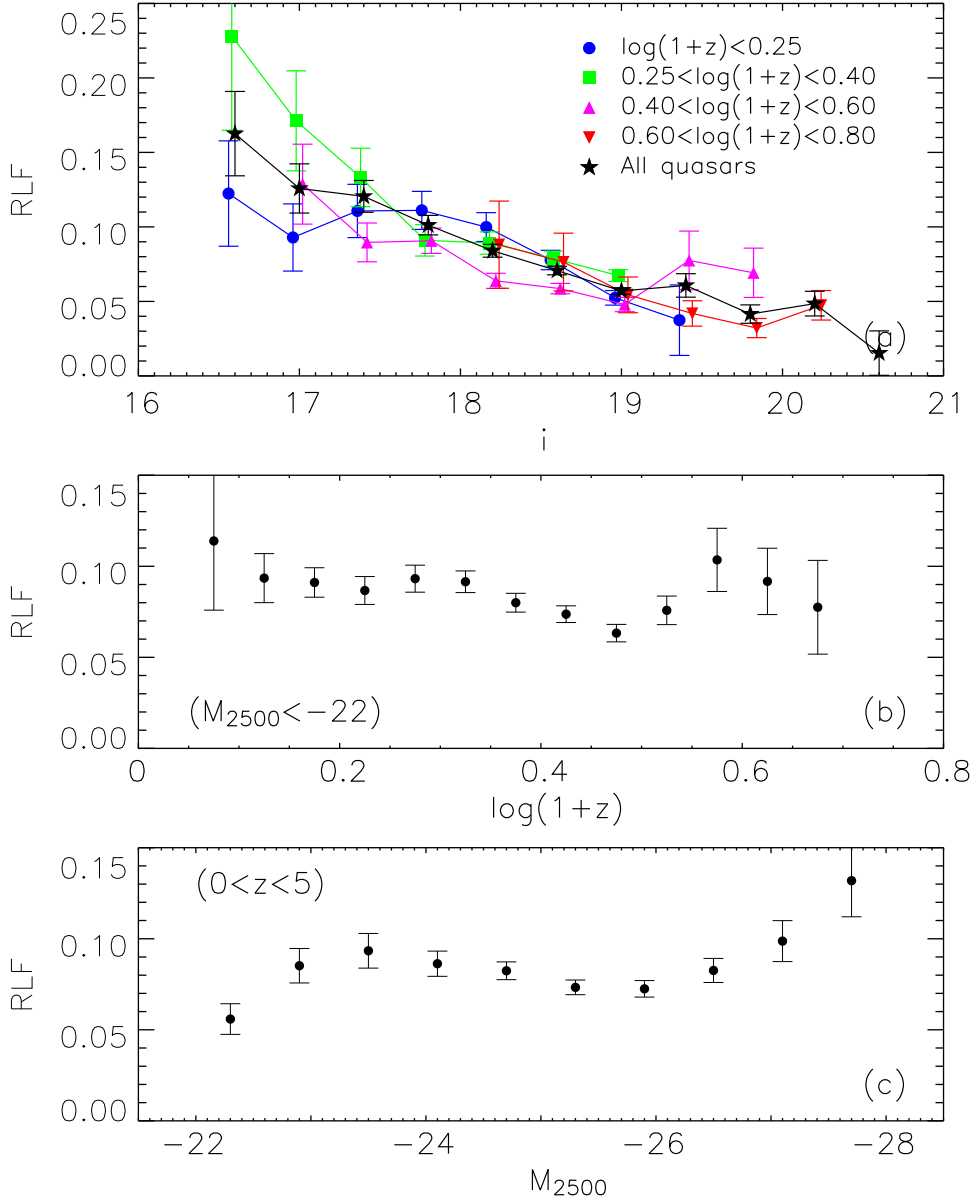}
\caption{(a): RLF as a function of apparent magnitude $i$ for five redshift
bins. The RLF declines with increasing $i$, and the curves for different
redshift bins follow the same RLF--$i$ dependence. (b) and (c): Marginal RLF
as a function of redshift and luminosity for our sample. The marginal RLF is 
roughly independent of both redshift and luminosity due to the $M_{2500}$--$z$ 
degeneracy.}
\end{figure}

\clearpage
\begin{figure}
\plotone{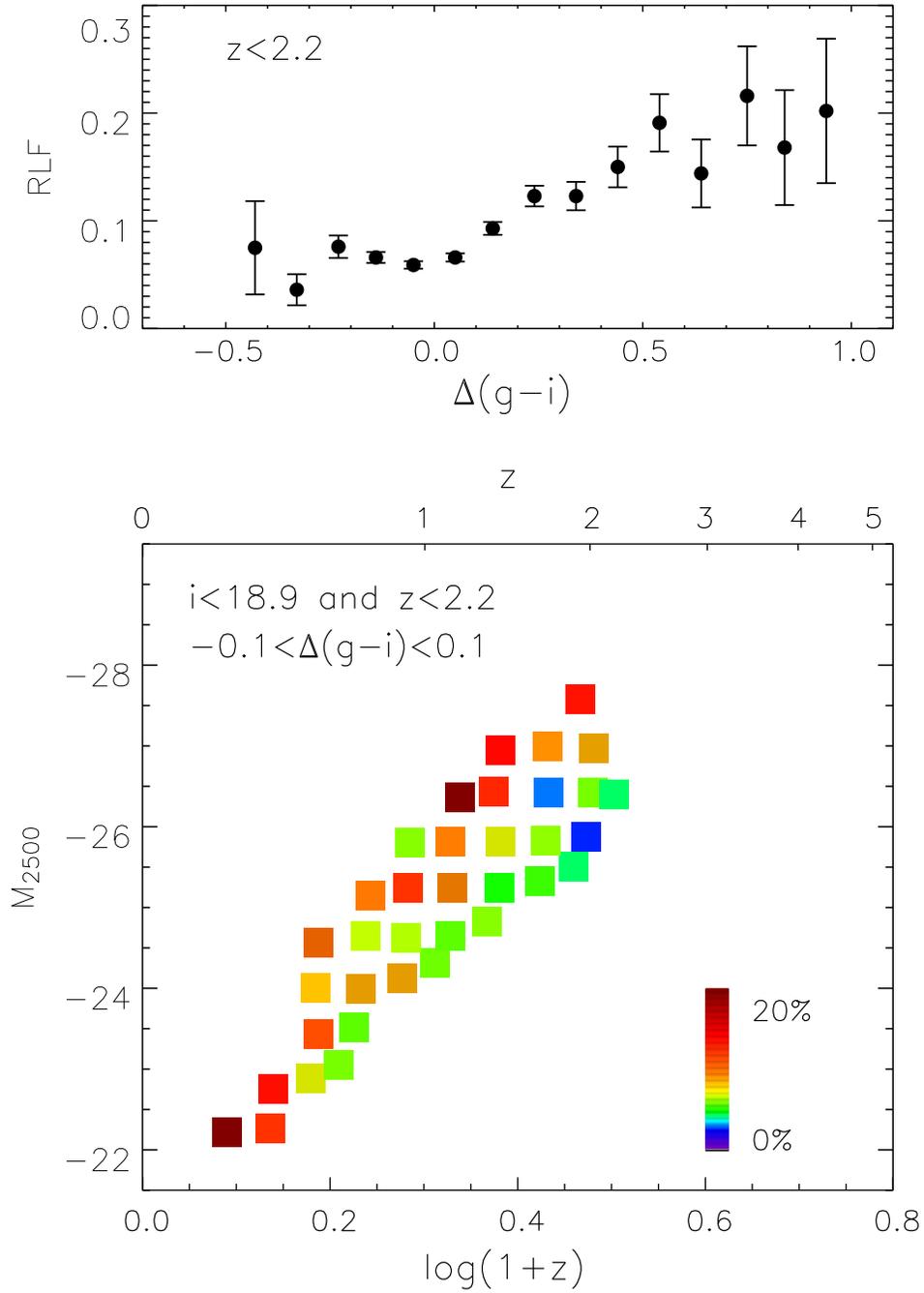}
\caption{Upper panel: RLF as a function of $\Delta(g-i)$. The RLF rises with 
increasing $\Delta(g-i)$ at $z<2.2$. Lower panel: RLF as a function of $z$ 
and $M_{2500}$ in a small color range of $-0.1<\Delta(g-i)<0.1$.}
\end{figure}
\end{document}